\documentclass[10pt,conference]{IEEEtran}
\usepackage{amssymb}
\usepackage{amsmath}
\usepackage{graphicx}
\usepackage{bm}
\usepackage{bbm}
\usepackage{subfigure}
\usepackage{psfrag}
\usepackage{pstricks, pstricks-add}
\usepackage{colortbl}
\usepackage{mathrsfs}
\usepackage{setspace}

\title{A Subsequence-Histogram Method for Generic Vocabulary Recognition over Deletion Channels}
\author{Majid Fozunbal\\
        Hewlett-Packard Laboratories\\
        Palo Alto, CA 94304\\
        \small{majid.fozunbal@hp.com}}

\newtheorem{example}{Example}[{section}] %
\def\a{\alpha}

\def\th{\theta}

\def\l{\lambda}

\def\s{\sigma}

\def\f{\phi}

\def\ps{\psi}

\def\S{\Sigma}

\def\mcal{\mathcal}

\DeclareMathAlphabet{\mathpzc}{OT1}{pzc}{m}{it}

\newcommand{\ccg}{\cellcolor[gray]{0}\color{white}}

\newcommand{\rcl}{\rowcolor[gray]{.8}}

\newcommand{\smp}[1]{P({#1})}
\newcommand{\vocab}[1]{\mcal V_{#1}}
\newcommand{\vosize}[1]{N_{#1}}
\newcommand{\indfun}[1]{\mathbb{I}(#1)}
\newcommand{\sindfun}[1]{{\bm 1}(#1)}
\newcommand{\vocspec}[2]{\Phi_{#1}(#2)}
\newcommand{\weightspec}[2]{\Psi_{#1}(#2)}
\newcommand{\avespec}[2]{\bar{\Psi}_{#1}(#2)}
\newcommand{\ord}{L}

\begin{document}
\maketitle

    \begin{abstract}

        We consider the problem of recognizing a
        vocabulary--a collection of words (sequences) over
        a finite alphabet--from
        a potential subsequence of one of its words. We
        assume the given subsequence is received  through
        a deletion channel as a result of transmission of a random
        word from one of the two generic underlying vocabularies.
        An exact maximum a posterior (MAP) solution for  this problem
        counts the number of ways
        a given subsequence can be derived from
        particular subsets of candidate vocabularies,
        requiring exponential time or space.

        We present a polynomial approximation  algorithm
        for this problem. The algorithm makes no prior assumption
        about the rules and patterns governing the structure of
        vocabularies. Instead, through off-line processing of vocabularies,
        it extracts data regarding regularity patterns in the
        subsequences of each vocabulary.
        In the recognition phase, the algorithm just uses
        this data, called {\em subsequence-histogram}, to decide
        in favor of one of the vocabularies. We provide
        examples to demonstrate the performance of the algorithm
        and show that it can achieve the same performance as MAP
        in some situations.

        Potential applications
        include bioinformatics, storage systems, and
        search engines.\\[-3pt]

    \end{abstract}

    \section{Introduction}
    \label{sec:intro}

        Consider you have two database of sequences.
        You observe  a subsequence that you  know
        is derived from one of the databases and you would like to determine which one.
        Alternatively, consider you have
        two nonlinear codebooks that you do not know their generating rules.
        Suppose a codeword is chosen from one of the
        codebooks, punctured with an unknown  deletion  pattern, and
        handed to you. Your task is to infer which codebook was chosen.
        As another example, suppose you have the vocabularies  of two languages: Spanish
        and Italian.  You see an abbreviated
        version of a word in one of them and try to decide which language
        it belongs to.

        We call these applications as generic {\em vocabulary recognition},
        in which one seeks to identify an underlying {\em vocabulary} (collection)
        based on a potential subsequence of one of its words.
        What make these problems particularly challenging are: 1)
        vocabularies are generic and 2) deletion patterns are unknown.
        Since vocabularies are generic, we do not know a
        common underlying model or  structure to simplify
        their descriptions. For example, should we know that the
        words in each vocabulary were generated by an i.i.d. source, then
        a simple {\em regular histogram} method could be used to learn
        the underlying distribution for each vocabulary and contrast it
        against that of the received subsequence. Moreover,
        the existence  of an unknown  deletion pattern makes the problem
        way more complex. Should there exist no deletion channel,
        then a {\em suffix-tree}  implementation of vocabularies
        would suffice to deal with the problem \cite{Algorithms:Strings:Gusfield}.
        In the existence of deletion, however,
        one needs exponential space to develop a generalized suffix-tree
        to enclose subsequences derived from a vocabulary.

        Deletion channels, not to be mistaken for erasure channels,
        have risen  in different applications. In a recent survey,
        Mitzenmacher \cite{DeletionChannels:Survey:Mitzenmacher} points out
        some of these applications as well as the existing open problems,
        including the capacity of the binary deletion channels.
        In \cite{Deletion:Diggavi:BufferChannels},
        Diggavi et. al study  communication  over a finite buffer in the framework  of
        deletion channels, and in \cite{DelectionChannels:Diggavi:Mitzenmacher}, they
        provide upper bounds for the capacity of these channels. As noted by these authors,
        the main  challenge is raised  because of the unknown deletion patterns.
        As a result, a maximum a posterior (MAP) or a maximum likelihood (ML) detector needs
        to  find the transmitted codeword analyzing  its subsequences.

        Problems involving subsequences are generally hard combinatorial problems
        whose exact solutions require    exponential time or space
        \cite{Subsequence:Complexity}, \cite{levenstein:Sequences}.
        This is no exception to decoding over deletion channels \cite{Deletion:Diggavi:BufferChannels} and
        to our case, in which, we formulate the vocabulary
        recognition problem in an MAP decision framework leading to the same
        bottleneck as in decoding: count the number of ways
        that a received subsequence can be obtained from a particular sequence.
        Unlike the case for decoding, here, we also need to compute
        the aggregated number of ways that a received subsequence can be derived
        from certain groups of words.

        Since the computation is NP-Hard, we approximate it
        by the average number of position-wise matches that a received
        subsequence can find in the multiset of same length subsequences
        derived from a vocabulary. This
        approximation led to an algorithm that we call
        {\em subsequence-histogram} method, due to its
        operational similarities to regular histogram method.
        In fact,
        one may view regular histogram method as an special case
        of subsequence-histogram method for subsequences of unit length.
        The algorithm requires a space
        proportional to the cube of the maximum length of words and
        a time that is upper bounded by the product of the length of the
        received subsequence and the maximum length of the words.
        Examples are provided to illustrate the process and performance
        of the algorithm. In one simple example, we show that the algorithm
        can achieve the same performance as the exact MAP.
        Another example, simulated numerically, shows a case
        in which the algorithm achieves 10\% error rate
        compared to the regular histogram method that has 50\%
        error rate. A rigorous performance analysis of the algorithm
        and assessment of its proximity to exact MAP is still
        work in progress.

        The organization of this paper is as follow. In Section \ref{sec:setup},
        we discuss the setup of the problem. In Section~III, we introduce the
        algorithm and provide examples to demonstrate its operation
        and performance. Finally, we will discuss the derivation of the algorithm and
        its error rate analysis in Section~IV.

    \section{Setup}
    \label{sec:setup}
        Let $n$ denote a discrete time index, and let $\S$ be a finite  alphabet.
        Assume two finite vocabularies $\vocab{\th_1}, \vocab{\th_2} \subset \cup_{l=1}^L\S^l$ of
        maximum order $\ord$ are given. A vocabulary is picked randomly with a
        known probability $\smp{\th}$ and a word $W=(w_1, \ldots, w_n)$ is picked from it uniformly random.
        The word $W$ is passed through a deletion channel producing
        $S=(s_1, \ldots, s_m)$ at the output. The channel is i.i.d.
        in which every letter $w_i$ of the word $W$ can be deleted with
        probability $p$. The problem is to observe $S$
        and infer which underlying vocabulary was more likely picked.
        \begin{example}
            Let $\S=\{0,1\}$ and assume \[\vocab{\th_1}=\{0101, 1100\}\] and \[\vocab{\th_2}=\{1010, 0011\}\] are
            two equiprobable underlying vocabularies. If $S=01$ is observed, $\vocab{\th_2}$
            is chosen since the likelihood of
            $\vocab{\th_1}$ and $\vocab{\th_2}$ are $\frac{1}{4}p^2(1-p)^2$ and $\frac{5}{12} p^2(1-p)^2$, respectively.
            On the other hand, observing $S=10$, $\vocab{\th_1}$ is chosen. Intuitively, this is because
            $S=10$ and $S=01$ are common subsequences in $\vocab{\th_1}$ and $\vocab{\th_2}$, respectively.
            \label{examp:MAP:elelementary}
        \end{example}

        The MAP formulation of  this problem is
        to maximize
        \begin{align}
            \smp{\th|S} &\propto \sum_{W\in \mcal V_\th} \smp{S|W} \smp{W|\th} \smp{\th},
            \label{eqn:map}
        \end{align}
        in which \[\smp{W|\th}=\frac{1}{\vosize{\th}}\]
        for every $W \in \vocab{\th}$.
         The challenge in Eq. \eqref {eqn:map} is to compute
        or approximate $\smp{S|W}$. Should there be no deletion, i.e.,
        $p=0$, then the observed sequence $S$ should exactly match
        a word $W$ in either of the dictionaries. In this case,
        a suffix-tree implementation of the vocabularies would require
        $O(|\vocab{\th}| L)$ space and allow to identify an
        exact match of sequence $S$ in $O(m)$ time. For $p\neq 0$, a
        generalized suffix-tree is not an option as it needs to enclose
        the multi-set of all possible subsequences of $\vocab{\th}$,
        requiring $O(|\vocab{\th}| 2^L)$ space, in the worst case.


    \section{Algorithm}
        Here, we introduce an approximation algorithm
        that needs $O(|\S|L^3)$ space and operates in
        $O(mL)$ time. The algorithm has two phases: a learning phase and
        a recognition phase.

        \subsubsection*{Learning Phase}
        In the learning phase, it computes two sets of matrices.
        The first set of matrices are what we call as
        {\em positional histograms}. That is for every
        $n \leq L$, we compute $|\S| \times n$
        matrices $\Phi_n$ whose elements are
        \begin{eqnarray}
            \vocspec{n}{\s, j} = \sum_{W\in \vocab{\th}(n)} \sindfun{w_j=\s}
            \label{eqn:phis}
        \end{eqnarray}
        for $\s \in \S$ and $1\leq j\leq n$. Here,
        \[\vocab{\th}(n)=\{W\in \vocab{\th}: |W| =n\}\]
        denotes the subset of $\vocab{\th}$
        containing all words of length  $n$. Eq. \eqref{eqn:phis}
        denotes the number of $n$-length words whose $j$-th position (from left) is $\s$.
        Through an off-line process, one can adaptively compute
        $\vocspec{n}{\s, j}$ for every $\s$ and $j$ and store it
        in a $|\S|\times n$ matrix $\Phi_n$. The total space
        required for this purpose is $O(|\S|L^2)$. This process is
        summarized in Table \ref{tbl:learning:model}.

            \begin{table}[t]
                \centering
                \caption{Illustration of the learning phase of algorithm.}
                \label{tbl:learning:model}
                \begin{tabular}{|l|}
                    \hline
                    {$\qquad\qquad\qquad$  \sc I. Computing $\Phi_n$}\\
                    \hline

                    1. Initialization:
                        \\$\quad$- For $n\leq L$, $\s \in \S$, and
                    $j\leq n$,\\ $\qquad\qquad\qquad$ $\Phi_n(\s,j)=0$.\\[10pt]

                    2. Recursion: \\
                    $\quad$ - For $W \in \vocab{\th}$, let $n=|W|$.
                        For  any $j \leq n$,\\ $\qquad\qquad\qquad$ $\Phi_n(w_j,j)=\Phi_n(w_j,j)+1$.\\[10pt]


                    3. Termination: For  $n\leq L$, \\
                           $\qquad\qquad$  output $|\S| \times n$ matrix $\Phi_n$.\\[10pt]


                    \hline
                    {$\qquad\qquad\qquad$  \sc II. Computing $\Psi_{n,m}$}\\
                    \hline
                    1. Recursion: \\
                        $\quad$-For $n\leq L$, $m\leq n$, $i\leq m$, and
                        $\s \in \S$, \\ $\qquad\qquad$ $\weightspec{n,m}{\s, i}=\sum_{j=1}^n  \vocspec{n}{\s,j} \a_{n,m}(j,i)$.\\[10pt]

                    3. Termination: For  $n\leq L$ and $m\leq n$, \\
                              $\qquad \qquad$  output $|\S| \times m$ matrix $\Psi_{n,m}$.\\
                    \hline
                \end{tabular}

            \vspace{-10pt}
            \end{table}

        \begin{example}[Positional histograms]
            Consider the two given vocabularies in Example \ref{examp:MAP:elelementary}. Then,
            for $n=4$,
            \begin{eqnarray*}
                \Phi_{n, \th_1} = \left[
                            \begin{array}{cccc}
                            1 & 0 & 2 & 1 \\
                            1 & 2 & 0 & 1 \\
                            \end{array}
                        \right],\quad
                \Phi_{n, \th_2} = \left[
                            \begin{array}{cccc}
                            1 & 2 & 0 & 1 \\
                            1 & 0 & 2 & 1 \\
                            \end{array}
                        \right].
            \end{eqnarray*}
            For $n=1,2,3$, the positional histograms are zero.
            \label{examp:Phi}
        \end{example}

        The second set of matrices are what we call as {\em subsequence  histograms}.
        For every $n\leq L$, and $m \leq n$, we compute $|\S| \times m$
        matrices $\Psi_{n,m}$ whose elements
        are
        \begin{align}
            \label{eqn:psis}       \weightspec{n,m}{\s, i} &= \sum_{j=1}^n  \vocspec{n}{\s,j} \a_{n,m}(j,i)
        \end{align}
        where
        \begin{eqnarray}
               \a_{n,m}(j,i)=\binom{j-1}{i-1} \binom{n-j}{m-i}
               \label{eqn:alphanm}
        \end{eqnarray}
        for $\s\in \S$ and $1 \leq i\leq m$ denote the number of ways that an
        $m$-length subsequence can be derived from $\vocab{\th}(n)$
        such that its $i$-th  element is $\s$.
        Each matrix $\Psi_{n,m}$ may be viewed as
        positional histogram of the multiset of all $m$-subsequences of
        $\vocab{\th}(n)$. These matrices can be computed off-line and
        stored in $O(|\S| L^3)$ space. Table \ref{tbl:learning:model}
        illustrates this process.

            \begin{example}[Subsequence histograms]
                For the vocabulary $\vocab{\th_1}$ in Example \ref{examp:MAP:elelementary},
                the subsequence  histograms are
                \begin{align*}
                    \Psi_{4,1} = \left[
                                     \begin{array}{c}
                                        4  \\
                                        4  \\
                                     \end{array}
                                \right],\quad
                    &\Psi_{4,2} = \left[
                                     \begin{array}{cc}
                                        5 & 7 \\
                                        7 & 5 \\
                                     \end{array}
                                \right],\quad
                    \Psi_{4,3} = \left[
                                    \begin{array}{ccc}
                                    3 & 4 & 5 \\
                                    5 & 4 & 3 \\
                                    \end{array}
                                \right],\quad\\
                    &\Psi_{4,4} = \left[
                                    \begin{array}{cccc}
                                    1 & 0 & 2 & 1 \\
                                    1 & 2 & 0 & 1\\
                                    \end{array}
                                \right].
                \end{align*}
                For vocabulary $\vocab{\th_2}$, on the other hand,
                they are
                \begin{align*}
                    \Psi_{4,1} = \left[
                                     \begin{array}{c}
                                        4  \\
                                        4  \\
                                     \end{array}
                                \right],\quad
                    &\Psi_{4,2} = \left[
                                     \begin{array}{cc}
                                        7 & 5 \\
                                        5 & 7 \\
                                     \end{array}
                                \right],\quad
                    \Psi_{4,3} = \left[
                                    \begin{array}{ccc}
                                    5 & 4 & 3 \\
                                    3 & 4 & 5 \\
                                    \end{array}
                                \right],\quad\\
                    &\Psi_{4,4} = \left[
                                    \begin{array}{cccc}
                                    1 & 2 & 0 & 1 \\
                                    1 & 0 & 2 & 1\\
                                    \end{array}
                                \right].
                \end{align*}
                Note that $\Psi_{4,1}$ is exactly the same data
                of the regular histogram. Thus, we may view regular histogram as
                an special case of subsequence-histogram for subsequences of unit length.
                Moreover,  since  $\Psi_{4,1}$ is the same for both vocabularies,
                the regular histogram method would be no better than tossing a fair coin.
                In contrast,  $\Psi_{4,2}$, $\Psi_{4,3}$, $\Psi_{4,4}$ reflect the differences
                between $\vocab{\th_1}$ and $\vocab{\th_2}$, a phenomenon that empowers
                the proposed algorithm enabling it to extract some regularity patterns of a vocabulary.
            \end{example}

        \subsubsection*{Recognition Phase}
            Observing $S=(s_1, \ldots, s_m)$,
            we compute {\em the subsequence  similarity score of order $n$}
            \begin{eqnarray}
                \avespec{n,m}{S}=\frac{1}{m}\sum_{i=1}^m \weightspec{n,m}{s_i, i}.
                \label{eqn:warping:similarity}
            \end{eqnarray}
            This score is an upper bound that is used as an approximation for the number of ways that
            $S$ can be derived from $\vocab{\th}(n)$. Computing this
            score for all $n\in \{m, \ldots, L\}$, we will obtain
            the {\em total similarity score}
            \begin{align}
                \tilde{J}_\th(S) = \frac{\smp{\th}}{\vosize{\th}}\sum_{n=m}^{L} \avespec{n,m}{S} p^{n}
                \label{eqn:dicounted:similarity}
            \end{align}
            and choose the vocabulary with maximum score.

            \begin{table}[t]
                \centering
                \caption{The recognition phase.}
                \label{tbl:recongnition}
                \begin{tabular}{|l|}
                    \hline
                    {$\qquad\qquad$\sc Computing Similarity Score}\\
                    \hline

                    1. Similarity score: Given $S=(s_1, \ldots, s_m)$, compute\\
                    $\qquad\qquad\qquad$ $\avespec{n,m}{S}=\frac{1}{m}\sum_{i=1}^m \weightspec{n,m}{s_i, i}$\\
                    $\quad$ for every vocabulary $\vocab{\th}$ and for $n\leq L$. \\[10pt]

                    2. Total similarity score:\\
                     $\qquad\qquad$   $\tilde{J}_\th(S) = \frac{\smp{\th}}{\vosize{\th}}\sum_{n=m}^{L} \avespec{n,m}{S} p^{n}$\\[10pt]

                    3. Recognition: $\vocab{\th}$ with maximum
                      $\tilde{J}_\th(S)$.\\[10pt]
                   \hline
                \end{tabular}
            \vspace{-10pt}
            \end{table}


                \begin{example}[Achieving MAP Performance]
                    Consider the same setup as in Example \ref{examp:MAP:elelementary}.
                    For an observed sequence $S$, the algorithm computes
                    the total similarity score \eqref{eqn:dicounted:similarity}
                    and makes the following decisions.
                    \begin{itemize}
                        \item $S \in \{0,1\} \rightarrow$ draw.
                        \item $S \in \{00, 11\}\rightarrow$ draw; $\;$ $S = 01\rightarrow \vocab{\th_2}$; $\;$ $S=10\rightarrow \vocab{\th_1}$.
                        \item $S\in \{100, 110\}\rightarrow\vocab{\th_1}$; $\;$ $S\in \{001, 011\}\rightarrow \vocab{\th_2}$;\\
                              $S \in \{010, 101\}\rightarrow$ draw.
                        \item $S\in \{0101, 1100\}\rightarrow\vocab{\th_1}$;$\;$ $S\in \{1010, 0011\}\rightarrow\vocab{\th_2}$.
                    \end{itemize}
                    \label{examp:discounted:similariy}
                \end{example}
                Should we have used exact MAP, the decision results were
                exactly the same as the algorithm's. Thus, this example shows
                that the algorithm can achieve the same performance as MAP. We,
                however, have no sufficient conditions for it, yet.
                One can also verify that the regular histogram method for the preceding example
                would be no better than tossing a coin.

                \begin{example}[All combinations]
                    Let $\S=\S_1\cup \S_2$ and $\S_1 \cap \S_2 \neq \varnothing$. Let $\vocab{\th_1}=\S_1^{L_1}$ and $\vocab{\th_2}=\S_2^{L_2}$.
                    Assume that the vocabularies are equiprobable and let $S \in (\S_1 \cap \S_2)^m$ be the observed sequence.
                    Then,
                    decision criterion would be to choose the vocabulary with maximum
                    \begin{eqnarray}
                        \tilde{J}_{\th_k}(S)=\frac{\binom{L_k}{m}  p^{L_k}}{|\S_k|}.
                    \end{eqnarray}
                    In other words, we the algorithm decides in favor of a vocabulary based on
                    \begin{eqnarray}
                        \frac{\binom{L_1}{m}}{\binom{L_2}{m}}  p^{L_1-L_2} \gtrless^{\th_1}_{\th_2} \frac{|\S_1|}{|\S_2|}.
                    \end{eqnarray}
                    Should have we used exact MAP, then 
                    recognition results would have been based on
                    \begin{eqnarray}
                        \frac{\binom{L_1}{m}}{\binom{L_2}{m}} p^{L_1-L_2} \gtrless^{\th_1}_{\th_2} \frac{|\S_1|^m}{|\S_2|^m}.
                    \end{eqnarray}
                    These two criteria only differ in the right hand sides.
                    Depending on the values of parameters, the two methods
                    may or may not have the same conclusion. In an
                    spacial case where $|\S_1|>|\S_2|$ and $L_1=L_2$, both methods
                    decide in favor of $\vocab{\th_2}$,
                    which has a smaller size alphabet. In this case,
                    regular histogram method has the same results as the
                    vocabularies have no specific pattern.
                \end{example}

                \begin{example}[i.i.d. sources]
                    Consider two i.i.d. sources over
                    alphabet $\S=\{\text{a, c, g, t}\}$. Let
                    \begin{align}
                        P_{\th_1} &= \left[\frac{1}{4}-\frac{i}{50}, \frac{1}{4}-\frac{i}{100}, \frac{1}{4}+\frac{i}{100}, \frac{1}{4}+\frac{i}{50}\right]\\
                        P_{\th_2} &= \left[\frac{1}{4}+\frac{i}{50}, \frac{1}{4}+\frac{i}{100}, \frac{1}{4}-\frac{i}{100}, \frac{1}{4}-\frac{i}{50}\right].
                    \end{align}
                    denote the density function of two i.i.d. sources on $\S$, in which parameter
                    $i \in \{0,1,2, 3,4\}$ is a deviation parameter. For each vocabulary and for each $i$, we use the
                    given densities to generate 128,000 random words of size 20 to 40.
                    A monte carlo error analysis, with 2000 trials, was then conducted.
                    Table \ref{tbl:error} summarizes the obtained error rate of the algorithm
                    for four different values of  the probability of deletion $p$. As
                    $i$ increases, the KL distance between the two distributions increases
                    and error rate decreases. In this example, regular  histogram method
                    has comparable error performance, as words are generated completely i.i.d..
                    \label{examp:iid:sources}
                \end{example}

                    \begin{table}[t]
                        \centering
                        \caption{Error results for i.i.d. sources. Each vocabulary contains
                        128000 words, generated as described in Example \ref{examp:iid:sources}.}
                        \begin{tabular}{|c|c|c|c|c|}
                                    \hline
                                        \ccg & \ccg $p = 0.1$ & \ccg $p=0.2$  & \ccg $p=0.3$ & \ccg $p=0.4$ \\
                                        $i =0$  & 0.50 & 0.50 & 0.50 & 0.49 \\
                                        \rcl $i =1$  & 0.37 & 0.38 & 0.39 & 0.40 \\
                                        $i =2$    & 0.25 & 0.27 & 0.27  & 0.29 \\
                                        \rcl $i =3$   & 0.16  & 0.18  & 0.19 & 0.21 \\
                                        $i =4$ & 0.09 & 0.10  & 0.12  & 0.14 \\
                                        \hline
                        \end{tabular}
                        \label{tbl:error}
                    \end{table}

                \begin{example}[Capturing inherent structures]
                    Using the same alphabet of the previous example,
                    two vocabularies are generated where the second one is the exact
                    horizontal mirror of the first one. The number of words are 128,000.
                    For the first vocabulary, the words
                    were created in five different  cases. In the first case, for each word,
                    a random length of size 20 to 40 is chosen and each letter of the word is chosen uniformly randomly from
                    $\S$. In the second to fifth cases, we simply add a prefix `a', `ac',
                    `acg', `acgt', and `acgta', to all words, respectively. In each case, we then
                    truncate from the end of the  words as many  letters as the length of the prefix added.
                    Table \ref{tbl:error_mirror} summarizes the results for
                    different cases and for four different probability of deletions.
                    Since the vocabularies are the exact mirror of each other,
                    regular histogram method resulted to $50\%$ error rate
                    in all cases. Note the big reduction in error rate
                    across all deletion probability
                    by adding just one letter of prefix. For example,
                    for deletion probability of $p=0.4$, the
                    error rate is reduced by half.
                    This shows that the algorithm is se successful
                    in capturing existing patterns in the vocabularies
                    and sensitive to words permutation.
                    \label{examp:mirror:sources}
                 \end{example}

    \section{Analysis of the algorithm}
            Maximum a posterior (MAP) method is equivalent to
            maximizing Eq. \eqref{eqn:map} in which the main
            challenge is to compute $\smp{S|W}$.
            Here, we discuss the derivation of an approximate
            algorithm for it.

            Assume a sequence $S$ of size $m=|S|$ is observed and let $W$
            be a word of length $n=|W|$. Sequence
            $S$ is  a {\em subsequence} of $W$,
            should there exists a {\em warping function}\footnote{A monotonically increasing sequence of indices.}
             \[\f_m=(i_1, \ldots, i_m), \; i_1<i_2<\ldots <i_m\leq n\]
            such that $S=\f_m(W)=(w_{i_1}, \ldots, w_{i_m})$. The function
            $\f_m$ may be viewed as an ordered $m$-subset in $[n]$. Thus,
            $S$ is a subsequence of $W$, if
            the number of such maps, i.e.,
            \begin{eqnarray}
                \ps(S,W)=|\{\f_m \subset [n]: S=\f_m(W)\}|
                \label{eqn:warping:number}
            \end{eqnarray}
            is non-zero. Using this expression, we can describe the
            probability of observing $S$ conditioned on $W$ as
            \begin{eqnarray}
                \smp{S|W}= \ps(S, W) p^{n-m} (1-p)^m.
                \label{eqn:S:cond:W}
            \end{eqnarray}
            Consequently, plugging \eqref{eqn:S:cond:W} in Eq. \eqref{eqn:map}, we obtain a discriminant
            function
            \begin{align}
                J_\th(S) = \frac{\smp{\th}}{\vosize{\th}} \sum_{n=m}^{L}p^{n} \sum_{W\in \vocab{\th}(n)} \ps(S,W).
                \label{eqn:map:plugin}
            \end{align}
            Exact computation of $\ps(S,W)$ requires dynamic programming that has
            $O(mn)$ time complexity. Thus, the computation of
            \eqref{eqn:map:plugin} is  $O(m\ord |\S|^\ord)$
            when  $|\mcal{V}_\th| = O(|\S|^\ord)$.

        \subsection{Approximation}
            We have
            \begin{align}
                \sum_{W\in \vocab{\th}(n)} \ps(S,W)  =\sum_{\f_m} \sum_{W\in \vocab{\th}(n)} \indfun{\f_m(W)=S}
                \label{eqn:sum:subdic:uval}
            \end{align}
            in which $\indfun{\cdot}$ is equality indicator function.
            Approximating $\indfun{\cdot}$  with the following
            upper bound
            \[\indfun{\f_m(W)=S}\leq \frac{1}{m} \sum_{i=1}^m \sindfun{\f_{m,i}(W)=s_i}\]
            and substituting in Eq. \eqref{eqn:sum:subdic:uval}, we obtain
            \begin{align}
                \nonumber \sum_{\f_m} \sum_{W\in \vocab{\th}(n)} \indfun{\f_m(W)=S}\leq& \\
                 \frac{1}{m}\sum_i \sum_{\f_m} \sum_{W\in \vocab{\th}(n)}&\sindfun{\f_{m,i}(W)=s_i}.
                \label{eqn:upperbound}
            \end{align}
            The right most summation returns the number of all
            $n$-length words whose $i$-th position under a warping $\f_m$ matches
            $s_i$. A  number
            \begin{eqnarray}
                \a_{n,m}(j,i)=\binom{j-1}{i-1} \binom{n-j}{m-i}
            \end{eqnarray}
            of theses warping functions
            map the $j$-th position of a word $W$ to the $i$-th position of
            the observed sequence $S$. This is the number of placement of
            $n$ distinguishable objects orderly into
            $m$ bins that place object $j$ into bin $i$
            satisfying the identity
            \begin{eqnarray}
                \binom{n}{m} = \sum_{j=1}^{n} \a_{n,m}(j,i)
            \end{eqnarray}
            for every $i=1,\ldots, m$.
            For any symbol $\s \in \S$, Eq. \eqref{eqn:phis}, i.e.,
            \begin{eqnarray*}
                \vocspec{n}{\s, j} = \sum_{W\in \vocab{\th}(n)} \sindfun{w_j=\s}
            \end{eqnarray*}
            denotes the number of words whose $j$-th position (from left) is $\s$.
            Eq. \ref{eqn:psis}, defined as
            \begin{eqnarray*}
                \weightspec{n,m}{\s, i} = \sum_{j=1}^n  \vocspec{n}{\s,j} \a_{n,m}(j,i)
                \label{eqn:psis:2}
            \end{eqnarray*}
            denotes the number of $m$-length subsequences, in the
            subsequence multiset derived from  $\vocab{\th}(n)$, whose $i$-th  element~is~$\s$.

            Using these
            notations, we can simplify the expression of right hand side of \eqref{eqn:upperbound} as
            \begin{align*}
                \frac{1}{m}\sum_{i=1}^m \sum_{\f_m} \sum_{W\in \vocab{\th}(n)}\sindfun{\f_{m,i}(W)=s_i}=
                    \frac{1}{m}\sum_{i=1}^m \weightspec{n,m}{s_i, i} ,
            \end{align*}
            which is the similarity score of $S$ and $\vocab{\th}(n)$
            measured as the average number of $m$-length subsequences that are derived from $\vocab{\th}(n)$
            and that match $S$ in different positions. This is Eq.~\eqref{eqn:warping:similarity}
            in the recognition phase that results to the total similarity score \eqref{eqn:dicounted:similarity}.


                   \begin{table}[t]
                        \centering
                        \caption{
                        Example \ref{examp:mirror:sources} demonstrating
                        that the algorithm is  capable of capturing inherent
                        structures in vocabularies and sensitive to words permutation.
                        }
                        \begin{tabular}{|c|c|c|c|c|}
                                    \hline
                                        \ccg & \ccg $p = 0.1$ & \ccg $p=0.2$  & \ccg $p=0.3$ & \ccg $p=0.4$ \\
                                        Case 1  & 0.49 & 0.50 & 0.49 & 0.49 \\
                                        \rcl Case 2  & 0.16 & 0.19 & 0.23 & 0.27 \\
                                        Case 3  & 0.08 & 0.13 & 0.19 & 0.23 \\
                                        \rcl Case 4   & 0.07 & 0.13 & 0.19  & 0.21 \\
                                        Case 5   & 0.09  & 0.14  & 0.19 & 0.23 \\
                                        \hline
                        \end{tabular}
                        \label{tbl:error_mirror}
                    \end{table}

        \subsection{Error analysis}
            Probability of error for this algorithm does not have a closed form solution.
            In an special case, in which vocabularies are equiprobable and
            have the same number of words, $N$, all of the same length, $n$, useful insights can  be obtained.
            In such case, observing $S$ of length $m$, the conditional probability of error
            for an exact MAP solution is
            \begin{align}
                P_{\text{MAP}}(\text{error})= \sum_{m=0}^{n}\frac{1}{2  N} p^{n-m}(1-p)^{m} \mu(m)
                \label{eqn:map:error}
             \end{align}
            where
            \begin{align}
                \mu(m) \triangleq \sum_{|S|=m}\min_{i}\sum_{W\in \vocab{\th_i}} \ps(S,W)
            \end{align}
            is the cardinality of the intersection of
            multisets of subsequences of length $m=|S|$
            derived from vocabularies. As $\mu(m)$
            increases, $P_{\text{MAP}}(\text{error})$ increases reaching a maximum
            of $\frac{1}{2}$ when the two vocabularies become identical.

            Expression for the error probability
            of sequence-histogram algorithm replaces $\mu(m)$ with
            \begin{align}
               \nonumber \l(m)= &\sum_{|S|=m}\sum_{W\in \vocab{\th_1}} \ps(S,W) \sindfun{\avespec{\th_1}{S}<\avespec{\th_2}{S}}\\
                            &+ \sum_{W\in \vocab{\th_2}} \ps(S,W) \sindfun{\avespec{\th_2}{S}<\avespec{\th_1}{S}}.
            \end{align}
            Using the inequality $\sindfun{x<1} \leq \frac{1}{\sqrt{x}}$, we will have
            \begin{align}
               \nonumber \l(m) \leq &\sum_{|S|=m} \sqrt{\avespec{\th_1}{S} \avespec{\th_2}{S}}\\
                    &
                    \underbrace{\left[\frac{\sum_{W\in \vocab{\th_1}} \ps(S,W)}{\avespec{\th_1}{S}} +\frac{\sum_{W\in \vocab{\th_2}} \ps(S,W)}{\avespec{\th_2}{S}}\right]}_{\leq 2}.
            \end{align}
            This implies the following upper bound on the probability of error of the algorithm:
            \begin{align}
               \nonumber P(\text{error}) \leq \sum_{m=0}^{n}&\frac{1}{N} p^{n-m}(1-p)^{m}\\& \sum_{|S|=m} \sqrt{\avespec{\th_1}{S} \avespec{\th_2}{S}}
            \end{align}
            Thus, we conclude that subsequences with
            equivalently large $\avespec{\th_1}{S}$ and $\avespec{\th_2}{S}$
            have bigger impact on error.

\vspace{10pt}
    \section{Conclusion}

        The aim of this paper is to demonstrate
        an approximation algorithm for the problem of generic vocabulary
        recognition over deletion channels. Without any prior assumption
        on the structure of vocabularies, the sequence-histogram algorithm
        seeks to extract regularity patterns of a vocabulary through an off-line
        analysis. The algorithm uses this data to choose the more
        likely underlying vocabulary for a received subsequence in
        polynomial time and space.

        A regular histogram method, may
        be viewed as an special case of this algorithm. Unlike a regular
        histogram, however, this algorithm is successful in extracting the
         structure of a vocabulary to dramatically boost its performance.
        In some situations, the algorithm can achieve the same
        performance as exact MAP. However, sufficient conditions
        to characterize such situations are not known, yet.

        Some immediate future directions are:
        1) analyzing the performance of the algorithm and its proximity to MAP,
        2) exploring applications in bioinformatics, sequence segmentation, storage systems, and search engines,
        3) extending work to  multiple observations, represented by subsequences of different words within one vocabulary, and
        4) generalizing model to include substitution and insertion  errors.

\vspace{10pt}
    \section*{Acknowledgement}
        The author is thankful to Krishna Viswanathan of HP Labs for
        his insights and  discussions on strings and deletion channels.
        The author is also grateful to the anonymous reviewers whose
        comments and suggestions shed new light on potential applications and
        future directions of this work.

\end{document}